\documentclass{dune}
\pdfoutput=1
\usepackage{mhchem}
\input{common/preamble}

\begin{document}

\pagestyle{titlepage}


\pagestyle{titlepage}

\newcommand{\Editor}{Editor.}
\newcommand{\Contributor}{Contributor.}

\date{}

\title{\scshape\Large Snowmass Neutrino Frontier: \\ Neutrino Interaction Cross Sections (NF06) \\ Topical Group Report\\
\vspace{5mm}
\normalsize Submitted to the Proceedings of the US Community\\
Study on the Future of Particle Physics (Snowmass 2021) \\
\vspace{10mm}
\vskip -10pt
}


\renewcommand\Authfont{\scshape\small}
\renewcommand\Affilfont{\itshape\footnotesize}

\newcommand*\samethanks[1][\value{footnote}]{\footnotemark[#1]}


\author[11]{A. B. Balantekin\thanks{Editor}}
\author[3]{S. Gardiner\samethanks}
\author[6]{K. Mahn\samethanks}
\author[3]{\\T. Mohayai\samethanks}
\author[8]{J. Newby\samethanks}
\author[3]{V. Pandey\samethanks}
\author[3]{J. Zettlemoyer\samethanks\vspace{0.3 cm}}


\affil[1]{University of Cincinnati}
\affil[2]{Deutsches Elektronen-Synchrotron DESY}
\affil[3]{Fermi National Accelerator Laboratory}
\affil[4]{Institut f{\"u}r Physik \& Exzellenzcluster PRISMA, Johannes Gutenberg-Universit{\"a}t Mainz}
\affil[5]{Iowa State University}
\affil[6]{Michigan State University}
\affil[7]{Nagoya University}
\affil[8]{Oak Ridge National Laboratory}
\affil[9]{University of Texas at Arlington}
\affil[10]{York University}
\affil[11]{University of Wisconsin, Madison}


\author[9]{\authorcr J. Asaadi\thanks{Contributor}}
\author[3]{M. Betancourt\samethanks}
\author[10]{D.~A.~Harris\samethanks}
\author[3]{A. Norrick\samethanks}
\author[2]{F.~Kling\samethanks}
\author[3]{B. Ramson\samethanks}
\author[5]{M.~C.~Sanchez\samethanks}
\author[7]{T.~Fukuda\samethanks}
\author[1]{M. Wallbank\samethanks}
\author[4]{M. Wurm\samethanks}

\maketitle

\renewcommand{\familydefault}{\sfdefault}
\renewcommand{\thepage}{\roman{page}}
\setcounter{page}{0}

\pagestyle{plain}
\clearpage
\textsf{\tableofcontents}

\renewcommand{\thepage}{\arabic{page}}
\setcounter{page}{1}

\pagestyle{fancy}

\fancyhead{}
\fancyhead[RO]{\textsf{\footnotesize \thepage}}
\fancyhead[LO]{\textsf{\footnotesize \rightmark}}

\fancyfoot{}
\fancyfoot[RO]{\textsf{\footnotesize Snowmass 2021}}
\fancyfoot[LO]{\textsf{\footnotesize NF06 Topical Group Report (2022)}}
\fancypagestyle{plain}{}

\renewcommand{\headrule}{\vspace{-4mm}\color[gray]{0.5}{\rule{\headwidth}{0.5pt}}}



\clearpage

\section{Executive Summary}
\label{sec:summary}

Neutrino interactions constitute either signal or background to a variety of exciting physics measurements planned in the near and far future. It is very important, for example, to model neutrino interaction cross sections robustly for the current and future program of neutrino oscillation measurements. Future programs recognize this, and so have designed highly capable suites of near detectors, with new capabilities to make cross section measurements needed for their programs; DUNE's program includes the use of precision detectors and different energy profiles\footnote{The energy profiles are achieved by ``off-axis'' detector positions relative to the beam.}.

There are valuable proposed cross section measurements for accelerator and atmospheric oscillation programs. These include new  electron scattering measurements (E12-14-012, e4nu, LDMX, A1, eALBA)
new pion scattering measurements (LArIAT, WCTE, ProtoDUNE), long-baseline measurements (T2K, NOvA, DUNE, HK), short-baseline measurements (MicroBooNE, SBND, ICARUS) and dedicated neutrino scattering measurements (ANNIE, MINERvA, NINJA, H/D bubble chambers, LHC, nuSTORM). These measurements provide new information to develop interaction model theory and improve its implementation in \textit{event generators}, the computer programs used to simulate neutrino scattering in the context of experimental analyses. Support for all stages of the process, from making new measurements, to improving theoretical neutrino scattering models, to refining interaction simulations in event generators, will be necessary to meet the needs of the precision experimental program. These efforts require, and benefit from, significant expertise from both High Energy Physics (HEP) and Nuclear Physics (NP) communities. Bringing together HEP and NP communities interested in neutrino cross sections for frequent exchange of ideas, results, techniques, and tools would pay significant dividends. Efforts for closer interaction between these two communities should be fostered.


The community faces important challenges:
\begin{itemize}
    \item The exact role of and impact of the suite of new cross section measurements on oscillation physics is not yet completely elucidated.
    A dedicated exercise, overseen by oscillation experimental programs but also involving theory and external measurements, is needed to assess the benefits of those measurements, and to refine what specific measurements could be valuable.
    \item  Event generators are a key tool for neutrino experiments. There are several available generators, each using different approaches, and all valuable to the community. However, improvements to these generators can take significant time to implement, and it is challenging to keep them up to date with state-of-the-art modelling. This is in part due to the necessary work being undervalued relative to other activities. The situation needs to be improved, and incentives aligned with the needs of the experimental program should be provided.  We also advocate for continued grassroots effort to identify and resolve short-term issues and to identify how generators should interface most constructively with experiments.
    \item There will be a wealth of important experimental cross-section data in the short term, but the needs of the oscillation program may change. We endorse efforts by experimental collaborations to have data preservation plans which allow for future re-analysis of the unique capabilities of experiments.
\end{itemize}




Low-energy neutrino scattering on nuclei will be a signal or background process in a variety of exciting physics measurements in the near and far future.
The recent observations of coherent elastic neutrino nuclear scattering (CEvNS) have spurred renewed interest in the process as a probe of fundamental properties of the neutrino, weak interactions, and nuclear properties. First-light measurements have already produced significant new constraints in these areas and the impact of future precision measurements of the total cross sections and recoil distributions on these areas as well as on searches for new physics is still being explored theoretically. This process will also be an irreducible background for direct dark-matter searches and a controllable background for accelerator-produced dark-matter searches at stopped-pion facilities where preservation of the timing structure of the neutrino flavors allows the measurement of the delayed neutrinos to provide a strong systematic control on the CEvNS background of the prompt neutrinos that largely overlap with the dark-matter signal.  

Low-energy inelastic scattering processes provide the foundational detection mechanism for solar, reactor, stopped-pion-based oscillation experiments and supernovae. In many cases, the cross sections for these interactions have never been measured, and the simulation capabilities needed to interpret future measurements are very limited. In some cases, there are nuclear physics topics interesting in their own right that can be explored in dedicated neutrino experiments.
\begin{itemize}
    \item Future planned CEvNS (COHERENT, CCM) measurements will make significant improvements in precision that expand our understanding of neutrino properties and extend the sensitivity of BSM tests and hidden-sector searches; electron-scattering experiments play a vital role in isolating modifications to the cross sections from the finite size of the nucleus.
    \item Measurements of inelastic cross sections on nuclei relevant for oscillation and supernova physics are critical to the success of these programs. While facilities and detection technologies exist to make these measurements, focused and dedicated efforts will be required to achieve the needed precision. Plans for such measurements on argon and oxygen below $\sim 50$ MeV are in the very early stages.
\end{itemize}


\section{Introduction}

A thorough understanding of neutrino cross sections in a wide range of energies is crucial for the successful execution of the entire neutrino physics program. In order to extract neutrino properties, long-baseline experiments need an accurate determination of neutrino cross sections within their detector(s). Since very few of the needed neutrino cross sections across the energy spectrum are directly measured, we emphasize the need for theoretical input and indirect measurements such as electron scattering, which would complement direct measurements. In this report we briefly summarize the current status of our knowledge of the neutrino cross sections and articulate needs of the experiments, ongoing and planned, at energies ranging from CEvNS and supernova neutrino energies to the DUNE and atmospheric neutrino energies. 

Since the last HEP long range planning activity, HEP experimenters successfully detected coherent elastic neutrino scattering from nuclei (CEvNS) for the first time \cite{Akimov:2017ade}. This was a most valuable addition to the existing handful of direct measurements of neutrino cross sections. At lower energies, relevant to CEvNS and supernova neutrino observations, neutrinos primarily interact with the neutrons in the target since their interaction with protons is suppressed by a factor of $(0.25 -\sin^2 \theta_W)$. This feature makes parity-violating electron scattering experiments provide very valuable input since the exchanged $Z$-boson at those experiments also interacts primarily with neutrons in the target due to the same suppression. At higher energies, relevant to experiments such as DUNE, theory input becomes even more important. 

At the very lowest energies well below the pion-production threshold, such as the break-up of deuteron by solar neutrinos (as was observed by the Sudbury Neutrino Observatory), one-parameter chiral effective field theory can be used to calculate the cross sections \cite{Butler:2000zp}. This parameter, describing the two-body isovector axial current, can be calculated using lattice gauge theory techniques \cite{Savage:2016kon}. As the neutrino energy increases and the targets needed become increasingly more complex nuclei, an extension of such an approach becomes more and more involved as both the parameters needed in the effective field theory description of the nucleon-nucleon interaction and the size of the Hilbert space needed to describe the target grow. For supernova neutrinos it is then necessary to use phenomenological techniques of the nuclear structure physics, such as the nuclear shell model. As the neutrino energy continues to increase and many inelastic and particle production channels open up, knowledge of the parton distribution functions in the target becomes crucial. 

In reactor neutrino experiments, the main channel is inverse beta decay, the cross section of which is reasonably well understood. Knowledge of the neutrino cross sections on $^{12}$C and $^{13}$C \cite{Fukugita:1988hg,Engel:1996zt,KARMEN:1998xmo,Kolbe:1999au,Hayes:1999ew,Volpe:2000zn,LSND:2001fbw,LSND:2002oco,Suzuki:2019cra} used in the scintillators also helps to assess subleading contributions. For supernova neutrino experiments using water Cerenkov detectors, inverse beta decay is still the dominant channel, with subdominant contributions coming from scattering on $^{16}$O nuclei. For experiments using liquid argon detectors there are significant challenges since neutrino-$^{40}$Ar cross sections need to be available for the wide energy range required. Additional information about semi-exclusive processes, such as neutron emission, is also crucial. The fact that $^{40}$Ar is an open-shell nucleus compounds the theoretical difficulties. 

\section{What we do and don't know about neutrino cross sections}


In this section, we follow the conventions from Ref~\cite{Formaggio:2012cpf}.





\subsection{Threshold-less and Low-Energy Nuclear Processes: $E_\nu \sim$ 0-100 MeV}


CEvNS is a neutral- current process in which a neutrino elastically scatters off the whole nucleus. The first detection of CEvNS was by the COHERENT collaboration \cite{Akimov:2017ade} and opened an exciting chapter of using CEvNS to test not only the Standard Model but also search for new physics \cite{Patton:2012jr,Coloma:2017ncl,Liao:2017uzy,Cadeddu:2017etk,Farzan:2016wym,Farzan:2018gtr,Abdullah:2018ykz,Denton:2018xmq,Canas:2018rng,Esteban:2018ppq,AristizabalSierra:2018eqm,Billard:2018jnl,Dutta:2019eml,Dutta:2019nbn,Cadeddu:2020lky,Miranda:2020tif}. The cross section increases roughly as $N^2$, where $N$ is the number of neutrons in the target. This cross section is very sensitive to the neutron distributions in the nuclear targets, which dominate the theoretical uncertainties \cite{Payne:2019wvy,Hoferichter:2020osn,CONNIE:2019swq}. As CEvNS experiments continue to improve their experimental precision so that they can search for new physics, more precise knowledge of neutron form factors is needed. On the experimental side, this can be provided by parity-violating electron scattering experiments on the same targets. On the theoretical side, improvements in nuclear structure physics should lead to more precise calculations of the neutron distributions. Contributions from protons are suppressed, but as the CEvNS experiments get more and more precise, knowledge of proton distributions could also be needed, especially if the value of the neutrino magnetic moment turns out to be just below the current limits. 
\textbf{}
Inverse beta decay (IBD) is a relevant interaction mechanism in all supernova neutrino detectors using a target material that contains hydrogen. The cross section for this reaction is known to 1\% \cite{Strumia:2003zx}. Neutrinos can also inelastically scatter off nuclei via charged-current or neutral-current interactions in supernova detectors. For HK, detailed knowledge of neutrino-oxygen cross sections is needed~\cite{Haxton:1987kc}. DUNE will enable a high-statistics detection of supernova electron neutrinos~\cite{DUNEsupernova} via 
$\nu_e+\ce{^{40}Ar} \rightarrow e^- + \ce{^{40}K^{*}}$. This channel is also important for solar-neutrino studies, where impressive sensitivity may be possible~\cite{Capozzi2019}. To reconstruct the energy of the incoming neutrinos, we need a reliable model for the exclusive cross sections to various excited states in $^{40}$K. The energy transfer needed to create these excited states will be experimentally reconstructed by measuring nuclear de-excitation products, including $\gamma$-rays and, at slightly higher energies, nuclear fragments (neutrons, protons, deuterons, etc.). Undetected nuclear de-excitation products (particularly neutrons) can significantly bias tens-of-MeV-scale neutrino energy reconstruction~\cite{marleyPRC}, and thus reliably accounting for them is an important concern.

A full description of the neutrino-Ar scattering cross section is a formidable theoretical challenge for several reasons: the wide energy range required to fully analyse experimental data; the necessity to understand semi-exclusive processes, such as neutron emission; and the open-shell structure of $^{40}$Ar, which makes it challenging to accurately model the nucleus. At energies below 40~MeV, data from charge-exchange reactions and mirror nucleus beta decay can be used to constrain the leading contributions to the neutrino-argon cross sections. However, there are no direct measurements and no similar indirect experimental constraints above 40 MeV.

\subsection{Intermediate Energy Cross Sections: $E_\nu \sim 0.1-20$ GeV}


Neutrino cross sections are a key input to oscillation physics results, as described in a recent review~\cite{NuSTEC:2017hzk}\footnote{This article includes extensive definitions and details of all the relevant processes in this energy regime.}. Quoting from the theory white paper~\cite{Ruso:2022qes}:


\begin{quote}
    In general, for oscillation physics and rare or exotic searches, multiple processes contribute to signal selections.    In this energy regime, charged current quasi-elastic, multi-nucleon, resonant processes, deep-inelastic scattering, and transition region playing an increasingly important role for future oscillation measurements. The (anti)neutrino sources from accelerator-based on atmospheric neutrinos are broad spectrum in energy, so multiple channels contribute to event rates; the energy dependence of each process is important as oscillation depends on energy. But, rare charged or neutral current processes may also be important as signal or background as well, especially for exotics searches. For each process, well grounded theoretical predictions are needed to assess event rates and uncertainties. This is complicated by the nuclear dynamics of the target medium (commonly, carbon, oxygen or argon). Furthermore, neutrino experiments also need predictions for all relevant flavors of neutrinos ($\nu_e$, $\nu_\mu$, $\nu_\tau$) and antineutrinos, to perform appearance searches (e.g. $\nu_\mu \rightarrow \nu_e$) or for CPV measurements ( $\nu_\mu \rightarrow \nu_e$ vs. $\overline{\nu}_\mu \rightarrow \overline{\nu}_e$ ).   Furthermore, the signal selection  may depend on the composition and kinematics of exclusive final states. The unprecedented increases to beam exposure and detector size also enable explorations of final states in increasing detail.
\end{quote}


Neutrino oscillation experiments measure event rates:

\begin{equation}
N^{\alpha \to \beta} (E_{\text{reco}}) = \sum_i \phi_\alpha (E_{\text{true}}) \times \sigma^i_\beta(E_{\text{true}}) \times \epsilon^i_\beta(E_{\text{true}}) \times R_i (E_{\text{true}}; E_{\text{reco}}) \times P_{\alpha\beta}(E_{\text{true}}) 
\end{equation}
\label{eq:evrate}

where $\phi_\alpha$ is the neutrino flux of flavor $\alpha$, $\sigma^i_\beta$ is the cross section of flavor $\beta$ for interaction process $i$, $\epsilon_\beta^{i}$ is the detection efficiency, $P_{\alpha\beta}$ is the oscillation probability.
The flux, cross section and oscillation probability all depend on the neutrino energy
which depends on the neutrino energy, $E_{true}$. The cross section depends on more than just neutrino energy, including the reaction kinematics. 
The energy estimator ($E_{reco}$) uses only observable quantities, and the response function, $R_i$, relates the true quantities to the observables. 

As has been observed in the literature, mismodelling of the cross section can affect oscillation measurements, through the  $\sigma^i$ and $R_i$ factors which are based on a given model. Cross section modelling is also relevant to exotic and BSM searches~\cite{bsmwp} for the same reasons. Cross section measurements themselves may also be affected by cross-section modelling via the $\epsilon^i$ factor, as detection efficiency is estimated using a simulation with a particular interaction model choice.

Multiple energy measurements and/or kinematic variables are needed to assess a complete picture of interaction physics.  Since the last Snowmass process, there is a growing appreciation of the complexity of the problem that our current interaction models do not replicate nature, and this is reflected in the proposed design and capabilities of future experiments.  First, oscillation programs use event rates close to production at a ``near detector'' to test cross section models and reduce uncertainties:

\begin{equation}
N^{\alpha}_{ND} (E_{reco}) = \sum_i \phi_\alpha (E_{true}) \times \sigma^i_\alpha(E_{true}) \times \epsilon^i_\alpha(E_{true}) \times R_i (E_{true}; E_{reco}) 
\end{equation}
\label{eq:ndrate}

Near detectors have been in use from the first oscillation experiments. However, new experiments will include new features to improve our understanding significantly. DUNE, for example, plans a ``highly capable near detector complex'', described in Section~\ref{sec:nuscatt} which will be used to probe features of the interaction model in enormous detail.


Complementing the near detector programs are measurements of related processes, including  pion scattering and electron scattering, which measure specific parts of the cross section important to neutrino scattering\footnote{This is described in Sections~\ref{sec:piscatt} and \ref{sec:escatt}.}. 
Finally, dedicated neutrino-nucleus cross section measurements are useful to measure interactions on target materials of interest and/or new kinematics. These historically have been important, as the process to update interaction models is iterative with theory and takes time. Measurements made in advance of a program provide input to model improvements, and then new measurements, which supercede them, further refine and test the model and uncertainties.

Uncertainties in the interaction model are represented with a set of nuisance parameters in physics analyses (oscillation, exotic physics, cross section physics, etc.). These parameters may represent physical features of the interaction model, or may be empirical parameters representing open questions or issues in the model. The interaction parameters may be tuned\footnote{Tuning may also be done on the flux or detector models of an experiment.} to a variety of data--  if applicable, near detector data and external relevant data, e.g. electron scattering, pion scattering, or neutrino scattering, providing an improved estimate of the parameters with reduced uncertainties. We provide a historical example from the T2K experiment. In early oscillation analyses, the dipole axial mass, $M_A^{QE}$, was used to represent uncertainties in the nuclear models used at the time~\cite{T2K:2013bqz}. In subsequent versions of the oscillation analysis, this is treated as a physical parameter within the model (the single nucleon axial form factor mass for QE scattering) and nuclear degrees of freedom are accounted for by other parameters~\cite{T2K:2021xwb}.

The advantage of using a set of physical parameters
is to expose where the model is incomplete and to allow for comparisons with other data sets. But, in many cases, experiments (which start with models which are known to be incomplete) have to resort to empirical parameters to represent possible deficiencies in the models used, propagate data/simulation disagreements into physics results, or represent limitations of a particular model's software implementation. The need to interpret different data sets introduces still more complications --
generally, there are no ``pure'' probes of given physics effects, due to the convolution in Eq.~\ref{eq:evrate}.  In some physics programs, like searches for sterile neutrinos, the role of understanding a first principles model may be essential. One example is in Ref~\cite{Acero:2022wqg}:

\begin{quote}
One limitation of work to date on T2K is the completeness of the assessment of interaction model uncertainties as applied to short baseline analyses. T2K analyses so far assume no \numu disappearance, however the interaction model systematic uncertainties are assessed based on external and ND280 measurements. Those measurements are placed close to production and therefore could be sensitive to a \numu disappearance signal, potentially biasing a dedicated \numu disappearance search.  T2K studied the possible impact of a subset of interaction model uncertainties on a ND280 \numu disappearance result~\cite{Bordoni:2017zwi} and found it to be robust, but this does not consider a full re-assessment of where external data is used to inform the model. Current efforts in T2K cross-section measurements and the implementation of ab initio computations in the context of three-flavor analysis would greatly benefit such studies as well.
\end{quote}

This kind of subtlety motivates a strong theoretical understanding of neutrino interactions. There are also new, transformative, experimental developments underway. As described in Sections~\ref{sec:piscatt},~\ref{sec:escatt} and~\ref{sec:nuscatt}, new experimental ideas and older data sets are being brought into consideration, where some of the degeneracies in the interaction model can be broken.  

The ultimate impact of interaction model deficiencies can be only evaluated properly by a given experiment and is specific to the physics being studied in a particular analysis. Broadly, appearance searches, including CP violation searches, depend on robust predictions of  differences between neutrino and antineutrino cross sections and between flavors (especially $\nu_e$ and $\nu_\tau$ vs. $\nu_\mu$). Exotic signals may mimic conventional processes like NC$1\gamma$ production. However, while there is widespread agreement on the importance of understanding neutrino cross-sections, there is currently no set of clear priorities for dedicated cross section measurements nor theoretical developments that should be achieved in advance of the future program. This is a shared challenge, starting with the experimental programs themselves. The disadvantage of empirical parameters is that it can make it difficult for communities external to an experiment to assess the impact of various model improvements on the anticipated precision of an analysis. It is incumbent on the experimental community to define the set of requirements on features of the interaction model for a given physics analysis and provide a list of open problems faced by the experiment. This task can only be done completely by the experimental program, especially for oscillation analyses, where the near detector data does significantly reduce sensitivity to certain cross section modelling pathologies.

Once experimental groups provide general or specific targets, then the wider community
is in a position to identify and pursue measurements, theoretical developments, and improvements to simulations. This process is ideally done in a coordinated way\footnote{The pandemic has very much hurt efforts like this, which depend on different communities connecting and learning about each other.} to achieve consensus and disseminate lessons learned across experiments/physics results. 
The shared task of improving neutrino interaction modelling is iterative and takes time. Because the required expertise crosses NP and HEP experimental and theoretical communities, dedicated support for each community to participate is essential to success of future programs; for example, these challenges and possible solutions are discussed in Ref~\cite{Ankowski:2022thw} for the electron scattering community. As discussed in Section~\ref{sec:nuscatt}, there are new measurements with neutrino facilities of interest to nuclear physics communities, in addition to the benefits to oscillation and exotic physics programs from electron scattering data.





  


\subsection{High Energy Cross Sections: $E_\nu \sim 20$ GeV - 1 EeV}

Experiments like IceCube, KM3NeT are sensitive to very high energy neutrino interactions, which can be used to measure neutrino cross sections and/or set constraints on beyond the Standard Model physics, as the SM cross sections are typically well known in these regimes. 

Cross-sections may be determined by measuring neutrino absorption in the Earth as a function of neutrino energy and zenith angle.   These measurements become effective for energies above about a few TeV. For reference, for 40 TeV neutrinos, the path through the center of the Earth corresponds to about 1 absorption length.   IceCube has pioneered absorption measurements, studying both $\nu_\mu$  \cite{IceCube:2017roe} and all-flavor neutrinos \cite{IceCube:2020rnc}.  
Radio-detection experiments like the radio component of IceCube Gen2 \cite{IceCube-Gen2:2020qha} will use GZK neutrinos (or other high-energy sources) to extend these measurements to much higher energies, perhaps above $10^{20}$ eV.
These studies are sensitive to a variety of beyond-the-Standard-Model phenomena which could cause a large increase in the cross section \cite{Klein:2019nbu}.  
Measurements of inelasticity require direct observation of the neutrino interaction.  Inelasticity is of interest as a probe of a number of physics topics, including the $\nu:\overline\nu$ ratio, charm production in $\nu$ interactions, searches for dimuon production and tridents \cite{Zhou:2021xuh}, and of the $\nu_\tau$ flux.
To measure inelasticity with a natural beam, it is necessary to measure both the hadronic cascade and the lepton from an interaction.  IceCube has done this for $\nu_\mu$, measuring the neutrino inelasticity distribution at energies from 1 TeV to above 100 TeV.    
Inelasticity measurements nicely complement cross-section measurements.  If a new BSM  $\nu$ interaction is present that increases the cross section, there is no reason to expect it to have a similar inelasticity distribution to deep inelastic scattering; it should also be visible in the inelasticity distribution.  A similar comment applies to nuclear effects like shadowing or a colored glass condensate.  These phenomena will decrease the cross-section, and also alter the inelasticity distribution, since quarks with low Bjorken$-x$ correspond to high inelasticity interactions (but there is also $Q^2$ dependence) \cite{Klein:2020nuk}.


Based on recent and anticipated measurements, the primary limitation for future high-energy experiments will be statistical sample size and detector resolution rather than cross-section modelling.

At energies of 100's of GeV to a few TeV, neutrinos of all flavors are produced at the LHC in the far-forward region. As will be discussed in Sec.~\ref{subsubsec:fpf},  the first neutrino interaction cross-section will be measured in this energy region in upcoming Run 3 of the LHC and will significantly extend accelerator neutrino cross-section measurements. A dedicated Forward Physics Facility (FPF)~\cite{Anchordoqui:2021ghd, Feng:2022inv} is proposed to further exploit this region during the High-Luminosity LHC (HL-LHC) era.

\section{Hadron Scattering Measurements}\label{sec:piscatt}


One challenge in reconstructing neutrino interactions is the modeling of \textit{final-state interactions} (FSI), i.e., secondary hadronic interactions within the target nucleus. Misinterpreting the resulting event topology affects the resolution on measurements of the neutrino energy. Thin target measurements of hadronic interaction cross-sections can be used to validate FSI models (see for example Ref~\cite{PinzonGuerra:2018rju}). Here, pion, proton\footnote{Measurements of proton transparency measurements may also be made with electron scattering (Section~\ref{sec:escatt}).} or neutron beams can be used to measure cross-sections at momenta relevant to the models.\footnote{These measurements are also used to validate and set uncertainties on the reinteractions of pions in the detector model.} There are also important measurements made of pion and proton scattering for neutrino flux predictions, from NA61 and EMPHATIC, as described in the report from NF09 topical group.






There have been a few important new measurements in pion interactions. The Dual-Use Experiment at TRIUMF (DUET)~\cite{DUET:2015ybm} measured the pion-absorption and charge-exchange cross-section on carbon ~\cite{DUET:2016yrf} using a positively charged pion beam produced at TRIUMF with an momentum range from 200-300 MeV/$c$. Those measurements are relevant for scintillator-based cross section and oscillation programs. Measurements on argon are relevant to the DUNE and SBN programs. The Liquid Argon In a Testbeam (LArIAT) experiment~\cite{LArIAT:2019kzd} was a LArTPC that operated at the Fermilab Test Beam Facility collecting negatively-charged pion data at a range of energies from 100 to 700 MeV. LArIAT made the world's first pion-argon inclusive cross-section measurement~\cite{LArIAT:2021yix} and plans further measurements on $\pi^-$ capture at rest and $\pi^-$ absorption, looking at the outgoing proton multiplicity and kinematics.      ProtoDUNE-SP~\cite{DUNE:2021hwx, DUNE:2020cqd} is a 770-ton single phase LArTPC currently operating at the CERN Neutrino Platform as a prototype for the DUNE far detector modules. ProtoDUNE plans to make measurements of  $\pi^{+}$, proton, and $K^{+}$ reactions in the momentum range 0.3-7 GeV/$c$, including elastic, quasielastic, and inelastic (absorption and charge exchange) processes. ProtoDUNE also will be able to provide multiplicity and kinematic information for outgoing particles from these reactions. In addition to ProtoDUNE, the DUNE high-pressure gaseous argon TPC test beam will operate during the 2022/23 beam year with a cubic meter detector at $5$ bar in the tertiary charged-particle beamline at Fermilab that was formerly used for LArIAT. The detector aims to collect samples of low energy hadrons travelling through and interacting in the gas volume, both to characterise and commission the detector, and to make measurements of proton and pion scattering on argon.

The Water Cherenkov Test Experiment (WCTE) will operate at CERN and make measurements of pion scattering and secondary production of neutrons from interactions~\cite{ref:cern}. Measurements are anticipated to benefit programs like T2K and HK, and possibly IceCube.

The NOvA Test Beam has been collecting data with a scaled-down NOvA detector and new tertiary charged-particle beamline deployed at Fermilab since January 2020, and is expected to continue running until July 2022. The tagged particle samples collected already include several thousand protons, pions, and electrons ranging in energy from 0.4 GeV to 1.5 GeV. In addition to measurements meant to reduce detector response uncertainties, these data will be used to carry out dedicated measurements of proton and pion scattering in the detector to improve modeling of the final-state of neutrino interactions.

\section{Electron-scattering Measurements}\label{sec:escatt}

\begin{figure}
\begin{center}
\includegraphics[width=0.99\textwidth]{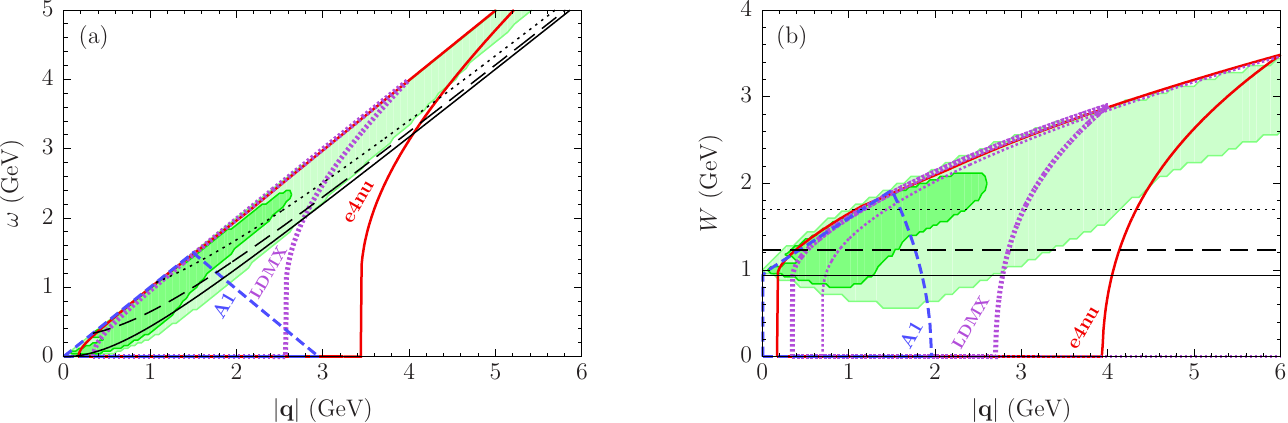}
\caption{Kinematic coverage of the ongoing and planned electron-scattering experiments on targets including argon and titanium, presented in the (a) $(|\mathbf{q}|, \omega)$ and (b) $(|\mathbf{q}|, W)$ planes, with momentum transfer $|\mathbf{q}|$, energy transfer $\omega$, and hadronic mass $W$. The light and dark shaded areas cover 68\% and 95\% of charged-current $\nu_\mu$Ar events expected in the DUNE near detector~\cite{DUNE:2021cuw}, according to GENIE 3.0.6. The thin solid, dashed, and dotted lines correspond to the kinematics of quasielastic scattering, $\Delta$ excitation, and the onset of deep-inelastic scattering at $W = 1.7$ GeV on free nucleons. Figure is taken from the white paper Ref.~\cite{Ankowski:2022thw}.}
\label{fig:e-scattering_future_coverage}
\end{center}
\end{figure}

 Electron-scattering provides a powerful probe of interaction models and their implementation in event generators~\cite{Ankowski:2019mfd}. In this case, particles are produced within the nucleus, so measurements are sensitive to in-medium effects. Furthermore, electron-scattering, which unlike a neutrino beam can be made to be monoenergetic, can be used to characterize a range of final states analogous to those found in neutrino scattering.
Current data is sparse, so measurements made in advance of the future program will enable us to probe kinematic regions for which our current understanding is poor. This is an important tool for assessing the viability of scattering models; and of evaluating the level of uncertainty introduced, when neutrino scattering is modeled with a theory that cannot fully reproduce nature. 
Electron-scattering measurements are complementary to those taken at near detectors in neutrino beams (Section~\ref{sec:nuscatt}); while electron-scattering can constrain the vector component of the neutrino-scattering cross section, near detectors are able to measure the axial-vector component. Both of these components are necessary for a complete model of the neutrino cross section. 
The ongoing and planned data for electron-scattering measurements is summarized in Table~\ref{tab:e-scattering_landscape} and the kinematic coverage is plotted on top of the DUNE coverage in Fig.~\ref{fig:e-scattering_future_coverage}, both taken from the white paper~\cite{Ankowski:2022thw}. The programs complement each other in kinematic reach and are at various stages of operation. Finally, as the neutrino and electron scattering programs continue to mature and interface, it may also be possible to take the electron scattering data and test it with targeted measurements in neutrino physics.  An example of such an exercise is described in Ref~\cite{Bodek:2018lmc},  applied to Ref~\cite{MINERvA:2019ope}.  Such work would maximize the important new advances in neutrino and electron scattering, including precision detectors, capabilities and unprecedented statistical power. 


\begin{table}
\centering
{\small
\begin{tabular}{ l | c c c c }
\hline {\bf Collaborations} & {\bf Kinematics} & {\bf Targets} & {\bf Scattering} \\
\hline
{\bf E12-14-012 (JLab)} & $E_e$ = 2.222 GeV & Ar, Ti & ($e,e'$) \\
{(Data collected: 2017)} & $15.5^\circ\leq{\theta_e}\leq21.5^\circ$ & Al, C & $e, p$ \\
 &$-50.0^\circ\leq{\theta_p}\leq-39.0^\circ$ &  & in the final-state \\
\hline
{\bf e4nu/CLAS (JLab)} & $E_e$ = 1, 2, 4, 6 GeV& H, D, He, & ($e,e'$) \\
{(Data collected: 1999, 2022)} & ${\theta_e} > 5^\circ$  & C, Ar, $^{40}$Ca, & $e, p, n, \pi,\gamma$ \\
 &  & $^{48}$Ca, Fe, Sn & in the final-state \\
\hline
{\bf LDMX (SLAC)} & $E_e$ = 4.0, 8.0 GeV & & ($e,e'$) \\
  {(Planned)} & ${\theta_e} < 40^\circ$ & W, Ti, Al & $e, p, n, \pi, \gamma$ \\
 &  &  &  in the final-state \\
\hline
{\bf A1 (MAMI)} & $50\text{ MeV}\lsim E_e \leq 1.5$ GeV & H, D, He & ($e,e'$) \\
 {(Data collected: 2020)} & $7^\circ\leq{\theta_e} \leq 160^\circ$ & C, O, Al & 2 additional \\
 {(More data planned)} &  & Ca, Ar, Xe & charged particles & \\
\hline
{\bf A1 (eALBA)} & $E_e$ = 500 MeV & C, CH & ($e,e'$) \\
 {(Planned)} & ~~~~~~- few GeV & Be, Ca & \\
\hline
\end{tabular}
}
\caption{Ongoing and planned electron-scattering experiments, table taken from the white paper Ref.~\cite{Ankowski:2022thw}.}
\label{tab:e-scattering_landscape}
\end{table}

\subsection{E12-14-012 experiment at JLAB} 
The E12-14-012 experiment at Jefferson Lab Hall A collected inclusive $(e,e^{'})$ (all electron-scattering events) and exclusive $(e,e^{'}p)$ (only events where a proton is ejected from the nucleus) electron-scattering data in the Spring of 2017 with five distinct kinematic setups using both a gas argon and a titanium target. The experiment used a 2.22~GeV electron beam provided by the Continuous Electron Beam Accelerator Facility (CEBAF) at Jefferson Lab. The data from the experiment determined the spectral function that represents the probability to remove a proton with momentum ${\bf p}$ from the target nucleus leaving the residual system with excitation energy $E-E_{\rm thr}$, with $E_{\rm thr}$ being the proton emission threshold. Final-state interaction effects should be accurately taken into account using an optical potential~\cite{giu87} for nuclei as heavy as $^{40}$Ar.

The experiment made a combination of inclusive and exclusive cross-section measurements. The inclusive cross-section has been measured on a variety of targets, including aluminum, carbon, and titanium and a closed argon-gas cell. Double-differential cross-section measurements over a broad range of energy transfer are reported with a high precision for targets all over the kinematic range~\cite{Dai:2018xhi,Dai:2018gch,Murphy:2019wed}. The exclusive cross-section has been measured on Ar with $\approx$4\% uncertainty using the data from the experiment. The cross-section was studied as a function of missing energy and missing momentum, and compared to the results of Monte Carlo simulations, and to the predictions of a model based on the Distorted Wave Impulse Approximation~\cite{osti_1770778}. The results of E12-14-012 will enable neutrino event generators to more accurately simulate the nuclear ground state of \isotope[40]{Ar} and thus better describe the interaction physics in liquid-argon-based detectors.

\subsection{$e4\nu$ at JLAB and CLAS12}
The $e4\nu$ effort at Jefferson Lab uses a large-acceptance detector to measure wide-phase-space exclusive and semi-exclusive electron-nucleus scattering at known beam energies from 0.5 to 12~GeV at the CEBAF facility. The goal of the experiment is to test energy reconstruction methods and interaction models.   Existing data has been taken with the CLAS spectrometer in 1999 and new data was taken in winter 2021/22 with the CLAS12 spectrometer.

The data taken with the CLAS spectrometer in 1999 includes $^3$He, $^4$He, C and Fe targets at 1.1, 2.2 and 4.4 GeV fixed beam energies. CLAS had the possibility to identify electrons, pions, protons, and photons, and to reconstruct their trajectories~\cite{CLAS:2003umf} 
covering a wide range of angles.
An e4nu enhancement to the GENIE event generator~\cite{PhysRevD.103.113003} enabled GENIE to simulate electron-scattering events, consistent with its neutrino event generation, and allowing for direct comparison~\cite{GENIEv3Highlights}.
Using the 1999 data with the CLAS spectrometer, cross-sections as a function of reconstructed energies were extracted for C and Fe at 1.159, 2.257, 4.453 GeV and compared to two different versions of GENIE~\cite{CLAS:2021neh}.

A dedicated $e4\nu$ measurement is being pursued using the upgraded CLAS12 spectrometer and data collected in winter 2021/2022~\cite{e4vproposal}. The CLAS12 spectrometer operates at a ten times higher luminosity than the previous CLAS spectrometer along with reduced scattering angle thresholds and neutron detection capabilities~\cite{BURKERT2020163419}. The experiment will use targets from D to Sn, including neutrino-detector materials (C, O, and Ar), at 1, 2, 4, and 6 GeV and other materials for calibration or nuclear physics purposes.

\subsection{LDMX at SLAC}
The Light Dark Matter eXperiment (LDMX) is a fixed-target electron-scattering measurement planned at SLAC to search for sub-GeV light dark matter. LDMX is synergistic with the planned US neutrino physics program with the requirement of precise reconstruction of both charged and neutral hadrons with the collection of a vast number of electron-scattering events~\cite{Ankowski:2019mfd}.

LDMX plans both inclusive and exclusive electron-scattering cross-section measurements. LDMX will collect a large amount of data at a beam energy of $4$~GeV, scattering angles below $40^\circ$, and energy transfers above $1$ GeV which covers a region of interest for the DUNE near detector~\cite{Ankowski:2019mfd}. LDMX is uniquely equipped to extract coincidence cross-sections for a variety of final-states involving pions and nucleons with a large sample of $10^{8}$ passing selection cuts of the $10^{14}$ electrons-on-target in the initial phase of the experiment.
This dataset will be sufficient to discriminate between, and to validate, the cross-section models used by different event generators.
If the target material currently used by LDMX were  varied in the future to a material such as high-pressure gaseous argon, helium, deuterium, or hydrogen the experiment would provide more data for nuclear modeling and understanding of the cross-section dependence on the atomic number. An extension of the selections to allow for energy transfers below 1~GeV would allow the experiment to fully cover regions where resonance currents and meson-exchange currents provide important contributions to the cross-sections.

\subsection{A1 Collaboration at MAMI}
The A1 collaboration hall at the Mainz Microtron (MAMI) facility focuses on electron-scattering experiments, and a dedicated experiment for neutrino scattering could be performed there. The current facility operates with a beam energy of 1.6 GeV with a maximum current of 100~$\mu$A. The A1 collaboration operates with a three-spectrometer setup designed to allow for the spectrometer to reach small angles. The collaboration routinely uses solid-state targets including C, Ca, Si, Ta, Pb. A cryogenic target is also available and has been in use with different elements in liquid phase (H, $^2$H, $^3$He, $^4$He) and its use for noble gases (e.g. Ar, Xe) is possible. A waterfall target is also available for measurements with oxygen and recently a supersonic gas-jet target was successfully tested with hydrogen and argon gases. 

\subsection{eALBA}
eALBA is a proposed multipurpose electron beam facility where the electron beam produced by the ALBA Synchrotron in Barcelona, Spain is extracted through the tunnel into an experimental hall. The planned infrastructure will cover a wide beam energy range from 100~MeV to 3~GeV with a low repetition rate (3~Hz) and high intensity per bunch (1~nC). One of the proposed experiments is intended to study electron-nucleus scattering with one possibility a atmospheric gaseous time-projection chamber (TPC) surrounding a target outside the gas volume. The experiment can explore different nuclei targets and energies using the planned external target. Beam energies spanning from $\sim$500~MeV to a few~GeV are needed to explore a range useful for neutrino experiments such as those using water Cherenkov technology. The measurements will cover several target nuclei including but not limited to C, CH, Be, and Ca. Gaseous and liquid targets including He, H$_2$O, and Ar are a challenge due to the size of the containment vessel.


\section{Neutrino Scattering Measurements}\label{sec:nuscatt}

Measurements of neutrino scattering are used to improve neutrino interaction models.
Oscillation programs, including short- and long-baseline programs, are both producers and users of cross-section measurements. In addition to measurements made at those facilities, there are also dedicated programs to understand neutrino interaction physics. The current landscape of neutrino scattering programs is summarized in Table~\ref{tab:expt}. Broadly, there is coverage of the kinds of experimental measurements needed for the future.  However, it would be beneficial to continue to assess sufficiency in the future, as some measurements are very challenging to make and the needs of the physics programs are refined.

\begin{table}[h]
\tabcolsep7.5pt
\caption{Current and future neutrino scattering experiments, broadly categorized by peak energy, and target material. T2K, HK, DUNE and MINERvA have multiple beam energies within the same beamline, due to either detector positioning, or beam configuration. In the case of MicroBooNE and ICARUS, neutrinos from the NuMI and BNB beamline provide two different energy spectra and flavor compositions. The PRISM technique involves detecting multiple distinct energy spectra in the same experiment, see text for details. 
}
\label{tab:expt}
\begin{center}
\begin{tabular}{@{}l|c|c|c|c@{}}
\hline
\hline
Experiment & Flavor &  $\nu_\mu$ Flux Peak (GeV) & Target & Detection \\
\hline
T2K & $\nu_\mu$,$\overline{\nu}_\mu$,$\nu_e$,$\overline{\nu}_e$ & 0.6,0.8,1 & CH, H$_2$O, Fe & Tracking \\
NOvA & $\nu_\mu$,$\overline{\nu}_\mu$,$\nu_e$, $\overline{\nu}_e$ & 2 & CH$_2$ & Tracking+Calorimetry \\
DUNE & $\nu_\mu$,$\overline{\nu}_\mu$,$\nu_e$,$\overline{\nu}_e$ & PRISM: 0.5-3  & H,C,Ar & Tracking+Calorimetry \\
HK IWCD & $\nu_\mu$,$\overline{\nu}_\mu$,$\nu_e$,$\overline{\nu}_e$ & PRISM: 0.4-1 & H$_2$O & Cherenkov \\
MicroBooNE & $\nu_\mu$,$\nu_e$ & 0.3,0.8  & Ar & Tracking+Calorimetry \\ 
SBND & $\nu_\mu$,$\nu_e$ & 0.8 (PRISM: 0.6-0.8)  & Ar & Tracking+Calorimetry \\ 
ICARUS & $\nu_\mu$,$\nu_e$  & 0.3,0.8  & Ar & Tracking+Calorimetry \\ 
MINERvA & $\nu_\mu$,$\overline{\nu}_\mu$,$\nu_e$,$\overline{\nu}_e$ & 3.5,6 & He, C, CH, & Tracking+Calorimetry \\
   &  &  & H$_2$O, Fe, Pb&  \\
ANNIE & $\nu_\mu$,$\overline{\nu}_\mu$ & 0.6  & CH, H$_2$O & Cherenkov \\ 
NINJA & $\nu_\mu$,$\overline{\nu}_\mu$,$\nu_e$,$\overline{\nu}_e$, & 1  & CH, H$_2$O, Fe & Emulsion \\
FPF & $\nu_\mu$,$\overline{\nu}_\mu$,$\nu_e$,$\overline{\nu}_e$, & 700 GeV &  W, Ar  & Emulsion, \\
 &  $\nu_\tau$, $\bar{\nu}_\tau$ &  &   &  Tracking+Calorimetry\\
nuSTORM & $\nu_\mu$,$\overline{\nu}_\mu$,$\nu_e$,$\overline{\nu}_e$ & PRISM: 0.8-3 &  CH,H$_2$O,Ar,TBD  & Tracking+Calorimetry (TBD)   \\
\hline
\hline
\end{tabular}
\end{center}
\end{table}

\subsection{Long-Baseline Experiment ND capabilities}

``Near'' detectors, situated close to the neutrino source, are used in long-baseline programs to  characterize the
energy spectrum and flavor composition of the neutrino beam prior to neutrino oscillation effects. The event rate at the near detector (if made of the same material as the far detector) may constrain aspects of the flux model, 
its response model, and the neutrino interaction model to reduce systematic uncertainties in the oscillation measurement. Near detectors make dedicated cross section measurements of processes of interest to oscillation, for iteration with theory. In addition, and crucially, near detectors also make cross section measurements for a wide variety of physics programs.  For example, near detectors make measurements of processes relevant to atmospheric neutrino measurements and proton decay. Near detectors also make measurements of nuclear physics (e.g. sin$^2\theta_W$, axial form factors).



\subsubsection{T2K-ND}



T2K is a long-baseline oscillation experiment taking data since 2009 with both neutrino and antineutrino enhanced beams. From the very beginning, T2K has been equipped of two near detectors: ND280 is placed 2.5 degrees off-axis, where the neutrino energy beam is peaked at  $\sim0.6$~GeV, while INGRID is placed on axis  where the neutrino energy beam is peaked at $\sim1$~GeV. Recently in 2019, a third near detector has been installed at 1 degree off-axis ($E_{\nu}^{peak}\sim0.8$~GeV) and is composed of a water and scintillator grid (WAGASCI) plus a magnetized range detector (BabyMIND). An upgrade of ND280~\cite{t2ktdr} is expected next year in view of the second phase of T2K. Processes relevant to oscillation physics programs~\cite{t2k1,t2k2,t2k3,t2k4,t2k5,t2k6,t2k7,t2k8,t2k9,t2k10,t2k11,t2k12,t2k13,t2k14} have been studied with the T2K near detectors, providing several cross-section measurements at different average neutrino energies: ND280 provided measurements reported in~\cite{t2k2,t2k3,t2k4,t2k6,t2k7,t2k10}; on-axis measurements are available in~\cite{t2k9,t2k11,t2k12} while the first WAGASCI-BabyMIND cross-section has been measured in~\cite{t2kref}.





In the near term, a final set of combined cross-section analyses using all T2K data and the latest reconstruction tools will supplement previous cross-section analyses using much of the T2K data. These analyses will include the incorporation of new systematics and test against the latest cross-section predictions and nuclear models. Possible analyses include the use of neutrino- and antineutrino-mode data, a water or hydrocarbon target, and multiple exclusive cross-section channels.
In the next several years, combined fits with samples taken from other detectors within the T2K beamline such as INGRID and WAGASCI can improve the accuracy of the measurements made with ND280 data alone.
A first analysis of this kind is nearing completion but additional data, especially from WAGASCI, along with a unified analysis framework, can allow T2K to constrain the neutrino flux and cross-section model.

On a longer-term path, the ND280 upgrade will provide several improvements to the overall T2K cross-section program. The planned upgrades enlarge the range of cross  section measurement capabilities in T2K thanks to a more efficient detection of the outgoing hadrons and to the increased lepton angle acceptance. In particular, ND280 upgrade will lower the detection threshold for protons and pions opening the door for more precise exclusive  measurements. On the other side, the increased angular acceptance for muons and pions allows a higher-statistics near detector sample, able to cover all regions of the allowed phase space.

Moreover, ND280 upgrade is expected to be able to measure neutrons using time-of-flight techniques, thus allowing an exclusive measurement of antineutrino interactions to very low neutron momentum thus improving our knowledge of this kind of interactions. This could help in constraining also neutrino interactions measurements (FSI, 2p2h).  Finally, the high electron reconstruction efficiency and purity expected with the upgrade will allow more precise $\nu_e$ and $\bar{\nu}_e$ cross-section measurements.

\subsubsection{NOvA-ND}

NOvA is planning to take data until 2025, collecting a large amount of statistics at the near detector which will enable a variety of measurements.
Muon-neutrino analyses will have the statistics to perform quadruple differential cross-section measurements in both the muon kinematics and the hadron system kinematics.
Similarly electron-neutrino analyses will perform quadruple differential measurements, with the total number of bins ($\approx 100$) limited by the resolution of the electron kinematic variables.
Charged-current analyses where a pion (charged or neutral) is selected, will report quadruple differential measurements of the lepton and pion kinematics, probing the resonant and soft inelastic scattering region.
Measurements of rare processes, like charged-current 
coherent scattering~\cite{nova-nd}, will report double or single differential results in the pion kinematics.
Finally, double differential (anti)neutrino electron-scattering measurements -- neutrino scattering from atomic electrons which has a lower cross section than neutrino-nucleus scattering, is free of nuclear or nucleon structure effects, and whose cross section is well-understood theoretically -- will reduce the flux normalization uncertainty from 10\% to a few percent, as all cross section uncertainty is removed from energy reconstruction calculations.



\subsubsection{DUNE-ND}
\label{sec:dunend}





 The DUNE ND~\cite{ref:DUNEwp} has been designed to support both the early oscillation physics program and to meet the needs of the ultimate exposure of DUNE, where systematic uncertainties on the interaction model are important.
DUNE's multi-megawatt intense (anti)neutrino source will also provide a wealth of ND data for unique cross-section measurements.

The DUNE ND complex will consist of a modular LArTPC (ND-LAr)\footnote{The ND-LAr detector must be used in conjunction with a downstream detector capable of reconstructing the momenta of muon tracks exiting ND-LAr in order to adequately match the muon acceptance of the FD. ND-GAr is one such detector which sits downstream of ND-LAr. In early running periods, a temporary muon spectrometer may be used before ND-GAr is placed in the beam.}, a pressurized gaseous argon TPC surrounded by a calorimeter and a magnet (ND-GAr), and a magnetized tracking spectrometer called system for on-axis neutrino detection (SAND).  ND-LAr and ND-GAr include the  same target material (Ar) as the far detector, to minimize effects of the nuclear model on the oscillation analysis. ND-LAr uses a similar technology to the FD, allowing for reduction of systematic uncertainties of the detector and interaction model simultaneously.




A pressurized gas argon TPC (HPgTPC) sits at the core of ND-GAr, surrounded by a calorimeter and magnet. HPgTPC has a lower detection threshold than ND-LAr and can reconstruct lower energy pion and proton tracks more effectively, thereby adding to our knowledge of the neutrino interaction constraints and hadronic kinematics~\cite{ref:ndgarwp}. It also has fewer secondary interactions such that the primary interactions can be easily distinguished from secondary interactions, thereby allowing to collect neutrino event samples that are less influenced by detector response and secondary interaction models~\cite{ref:ndgarwp}.
In addition, ND-GAr has full solid-angle acceptance and can complement ND-LAr by reconstructing particles that range out of ND-LAr, thereby adding back the missing ND-LAr kinematics phase space which would have otherwise led to uncertainties in the oscillation measurements. The HPgTPC portion of ND-GAr sits inside a magnet which will enable it to effectively distinguish between neutrinos and antineutrinos~\cite{ref:ndgarwp}.

PRISM is a key new capability of the DUNE ND complex~\cite{DUNE:2021tad} and other experiments.  As one moves laterally from the main axis of the neutrino beam
(off axis), the neutrino energy spectrum peak shifts. Data collected with the different spectra will include different combinations of interaction processes and can be combined to produce a data driven comparison to the FD oscillated spectrum or approximately monochromatic energy responses. ND-LAr and ND-GAr will move transversely to the neutrino beam axis, collecting neutrino flux data at various off-axis locations, while SAND will be kept fixed ``on axis'', to monitor the beam. This approach makes the oscillation analysis robust to interaction mismodellings.

The DUNE ND complex will also be able to make interesting new measurements relevant to nuclear physics. Different energy spectra enabled by the PRISM program can be combined to make  pseudo-monochromatic fluxes, analogous to electron scattering~\cite{DUNE:2021tad}. As this can be done for neutrinos and antineutrinos, DUNE can measure the axial-vector interference terms; for DUNE's energies, currently no measurements exist. This program is also enhanced significantly by  electron scattering measurements, which measure the vector response. 
SAND can provide cross-section constraints on multiple nuclear targets and can be used to make dedicated cross-section measurements, including sin$^2\theta_W$, tests of isospin physics and sum rules, and the strangeness content of the nucleon and measurements of neutrino-hydrogen scattering. In particular, by comparing cross-section measurements from its graphite (pure carbon) and plastic (CH$_2$) targets, SAND will be able to effectively isolate a cross section on (solid) hydrogen. As hydrogen consists of a single proton, this gives us opportunities to separate out nuclear effects from neutrino-nucleon scattering processes, allowing us to more effectively validate the overlapping components of complex neutrino-nucleus scattering models. 
HPgTPC portion of ND-GAr also has the flexibility to operate with a hydrogen-rich gas mixture as its nuclear target to enable measurements of neutrino-hydrogen interactions, giving direct access to fundamental physics parameters, such as proton radius and axial form factor~\cite{ref:ndgarwp}.

\subsubsection{HK-ND}

The ND for HyperKamiokande (HK) will include an on-axis beam monitor and magnetized off-axis detectors, with designs similar to those used on T2K.  There will also be an intermediate water Cherenkov detector (IWCD) located at $\sim 1$ km from production. This detector would move vertically to take data at a range of angles relative to the beam (PRISM effect).

\subsection{Short-Baseline Experiment ND capabilities}

The Short-Baseline Neutrino (SBN) program at Fermilab hosts a series of Liquid Argon Time Projection Chamber (LArTPC) detectors (SBND, MicroBooNE, and ICARUS) along the Booster Neutrino Beamline (BNB). MicroBooNE and ICARUS also have exposure to the off-axis NuMI beam. While the primary goal of the SBN program includes searches for exotic physics, like light sterile neutrinos, the program also provide extensive neutrino-argon cross section measurements for validation of theory, models and implementation in advance of DUNE operation.

\subsubsection{MicroBooNE}
The MicroBooNE detector recently completed operations in late 2021 after accumulating a data set of roughly five hundred thousand neutrino-argon scattering events. The first generation of cross-section measurements from MicroBooNE demonstrated the ability to reconstruct different final-state particles with low thresholds and to measure a variety of interaction channels~\cite{uboone1,uboone2,uboone3,uboone4,uboone5,uboone6,uboone7,uboone19}. Ongoing and future analyses will leverage greater statistics to measure cross-sections with higher dimensionality, separate interactions into more exclusive final-states, and search for rare processes. Meanwhile, MicroBooNE will continue to develop analysis methods for future argon-based efforts, such as LArTPC calorimetry, particle reconstruction, and novel cross-section extraction techniques including Wiener-SVD unfolding~\cite{uboone10,uboone11,uboone12,uboone13,uboone14,uboone15,uboone16,uboone17,uboone18,uboone20}.
Due to the long time-scale for iteratively improving neutrino event generators, current and near-future early measurements from MicroBooNE are helpful to guide event generator development. These data sets will be augmented by measurements at SBND and ICARUS in coming years.

\subsubsection{ICARUS (NuMI off-axis beamline)}





The ICARUS detector is located on-axis from the Booster neutrino beam and 103 mrad off-axis from the Neutrinos at the Main Injector (NuMI) beam. The ICARUS detector serves as the far detector of the Short Baseline Neutrino (SBN) program. 
 ICARUS will collect a rich data set for muon and electron neutrinos including quasi elastic, resonance, meson exchange current and deep inelastic scattering events from the NuMI neutrino beam. 
 ICARUS at the NuMI neutrino beam is scheduled to run for three years starting in early 2022. NuMI has a higher electron neutrino content than the BNB, dominated by  $K_+$ and $K_0$ decays, which can be used for measurements of the electron neutrino cross section. The majority of the neutrinos reaching ICARUS are produced near the beam target.

 ICARUS will make cross-section measurements, including the ratio $\nu_e/\nu_{\mu}$, inclusive and exclusive channels (resonance (RES) and deep inelastic(DIS)). RES and DIS are the dominant processes for DUNE\cite{tdr-vol-1,tdr-vol-2}. 
 The $\nu_e/\nu_{\mu}$ ratio uncertainty is a leading source of the cross-section uncertainty in the DUNE far detector\cite{tdr-vol-1,tdr-vol-2}. The $\nu_e$ and $\nu_{\mu}$ cross-section measurements from ICARUS will be beneficial to untangling the $\nu_e/\nu_{\mu}$ ratio uncertainty given its increased kinematic coverage and higher electron neutrino content.



\subsubsection{SBND}



The Short Baseline Near Detector (SBND), a 112 ton LArTPC detector, is the near detector (sited 110 m from the beam target) of the SBN program at Fermilab. With its proximity to the neutrino source, it will compile neutrino interaction data with an unprecedented high event rate and will provide an ideal venue to conduct precision studies of the physics of neutrino-argon interactions~\cite{Machado:2019oxb}. 
The experiment will make the world’s highest statistics cross-section measurements on argon, before DUNE, for both inclusive and many exclusive neutrino-argon scattering processes. SBND will perform many exclusive measurements of different final-states, utilizing LArTPC's capabilities, for $\nu_\mu$ and $\nu_e$ events with high precision. SBND data will allow the study of nuclear effects in neutrino interactions in argon nuclei with high precision. The high interaction rate will also allow SBND to measure several thousand events of many rare interaction channels, e.g., production of hyperons $\Lambda^{0}$ and $\Sigma^{+}$ in neutrino-argon interactions.

SBND’s large mass, proximity to a high intensity beam combined with high-statistics and LArTPC's capability enables it to exploit a ``PRISM''-like feature -- the decrease in the peak energy of the neutrino spectrum and the reduction in the size of the high energy tail when the detection angle relative to the neutrino beam axis is increased~\cite{nuPRISM:2014mzw, DUNE:2021tad}.
This feature allows additional degree of freedom in not only constraining neutrino interactions in the oscillation physics but also in performing targeted neutrino interaction measurements. By measuring neutrino interactions over a continuous range of off-axis angles from $0.2^{\circ}$ to $1.6^{\circ}$, therefore with different fluxes whose mean energy varies for $\sim$ 200 MeV, SBND-PRISM gives sensitivity to different parts of the neutrino cross-section. 


\subsection{Dedicated neutrino scattering programs}


\subsubsection{MINERvA}


MINERvA is a dedicated cross-section experiment at Fermilab, which took data from 2009-2019 with a variety of different target materials and beam energies~\cite{ref:1,ref:2,ref:3,ref:4,ref:5,ref:6,ref:7,ref:8,ref:9,ref:10,ref:11,ref:12,ref:13,ref:14,ref:15,ref:16,ref:17,ref:18,ref:19,ref:20,ref:21,ref:22,ref:23,ref:24,ref:25,ref:26,ref:27,ref:28,ref:29,ref:30,ref:31,ref:32,ref:33}. MINERvA has several mature analyses that will be released to the community through publications in the coming months. Some will be first-ever measurements of exclusive processes seen simultaneously across iron, lead, carbon, and water, all compared to scintillator. The collaboration is finalizing publications of both quasi-elastic and pion production measurements, where enough statistics have been recorded to do detailed investigation of the final-state particle kinematics. These measurements comprise the bulk of the processes that (will) occur at current (and future) accelerator-based neutrino oscillation experiments.  MINERvA is also able to measure and will soon publish the target-dependence of the charged current coherent pion cross-section, another first for the field. Finally, because of the fine granularity of the detector and the high statistics, MINERvA is working to finalize a measurement of the axial form factor of the proton, by using the kinematics of antinetrino scatters on our hydrocarbon target and using  data to predict and then subtract the background from interactions on carbon.  MINERvA is also nearing completion of a first ever comparison of electron neutrino and antineutrino interactions with low recoil.  These measurements are especially valuable to searches of electron neutrino appearance, including CP violation.
MINERvA is also finalizing analyses on neutrino and antineutrino interactions at higher momentum transfers all the way up to deep inelastic scattering, again with the goal of comparing the process on different nuclei.  The MINERvA data has far more capabilities than the modest size of the collaboration can realize, so they are also expending effort on preserving their data so that future analyzers in the community more broadly can investigate new ideas in neutrino interactions.

\subsubsection{ANNIE}

The Accelerator Neutrino Neutron Interaction Experiment~\cite{Back2020,annieloi} (ANNIE) consists of a 26-ton (2.5 ton fiducial) Gadolinium-doped, water Cherenkov based detector located in close proximity to to the Booster Neutrino Beamline (BNB) at Fermilab. The main physics thrust of ANNIE is to measure the multiplicity of final-state neutrons present in neutrino-nucleus interactions studying one of the dominant systematics in long-baseline neutrino experiments such as DUNE using novel LAPPD (Large Area Picosecond Photodetectors) technology. One possible extension to the ANNIE physics program is the measurement of quasi-elastic Neutral-current and charged-current interactions with low visible energy. ANNIE is located immediately upstream from SBND at the BNB so joint analyses are possible to constrain nuclear models through final-state multiplicity and kinematical measurements and also measurements of the H$_2$O/$^{40}$Ar cross-section ratio at GeV beam scales. Further upgrade paths include a WbLS option, enabling additional calometric information about the total hadronic activity which aids in event generator tuning. ANNIE also has the capability to use the bunch structure of the Fermilab beamline to study the relationship between neutrino beam composition and neutrino timing, which may be an interesting technique for long-baseline experiments such as DUNE.

\subsubsection{NINJA}

The Neutrino Interaction research with Nuclear emulsion and J-PARC Accelerator~\cite{ninjaNew1,ref:ninja1,ref:ninja2} (NINJA) experiment aims to measure sub-multi~GeV neutrino-nucleus cross-sections on water, CH and iron target precisely using nuclear emulsion films and a high-intensity neutrino beam from the J-PARC neutrino beamline.

Thanks to sub-micron spatial resolution and fine granularity of nuclear emulsion, it allows us to detect very short tracks of low-momentum charged particles, especially protons down to 200 MeV/c. These are useful to accurately study the 2-particle 2-hole (2p2h) neutrino interactions. These capabilities also allow separate measurements of electrons and positrons in pair-creation from gamma-rays, and provide good discrimination between electrons and gamma-rays, which is useful for suppressing background in electron neutrino CC interaction measurements. Therefore, the physics program of NINJA also includes measurements of electron neutrino cross-sections and searches for sterile neutrinos.

The data taking of emulsion films in the first physics run of NINJA in which it used a 250 kg (including 75 kg water, 130kg iron, 15kg plastic and 30 kg emulsion) nuclear emulsion detector, was completed this March with an automatic scanning system which can cover a wide-angle acceptance~\cite{ninjaNew2}. Behind the nuclear emulsion detectors is a scintillator detector dedicated to the NINJA~\cite{ninjaNew3}, and muon ID is performed by event matching with the Baby-MIND~\cite{ninjaNew4}, one of the T2K near detectors, using timing information. The analysis of the first physics run is underway and the second physics run is planned for the end of next year.
A NINJA-type water target nuclear emulsion detector is contemplated to be installed as a near detector for ESSnuSB~\cite{ninjaNew5}.

NINJA is also considering using a heavy water target instead of a water target to study neutrino-nucleon interactions by analyzing the subtraction between neutrino-heavy water and neutrino-water interactions. In fact, a test experiment using a nuclear emulsion detector for heavy water target was conducted in 2021~\cite{NinjaNew6}.

\subsubsection{H/D bubble chambers}



An independent nucleon-level neutrino amplitude measurement combined with precise nuclear models can be used to accurately measure the neutrino cross-sections~\cite{ref:hydrogenwp}. In addition, it is possible to significantly reduce cross-section uncertainties by factorizing nucleon level form factors from nuclear effects including meson exchange currents, final-state interactions, and initial state nucleon momentum distributions~\cite{ref:hydrogenwp}. Scattering measurements made on light nuclear targets such as hydrogen and deuterium are an excellent way to do this. In addition to the possible measurements discussed in Section~\ref{sec:dunend},
a dedicated facility can be built that uses a high-density liquid hydrogen/deuterium target to keep the proposed facility as small and cost-effective as possible~\cite{ref:hydrogenwp}. A modular bubble chamber design is being developed that combines many of the design principles of historic devices with engineering techniques from contemporary dark-matter focused chambers to produce a small modular device with flexible operation constraints and reduced cooling requirements (and costs)~\cite{ref:2ndhydrogenwp}.



\subsubsection{Far-Forward Neutrinos at the LHC}
\label{subsubsec:fpf}


The LHC produces an intense, strongly collimated and highly energetic (100's of GeV to a few TeV) beam of all flavors of neutrinos and antineutrinos in the far-forward region. These neutrinos are generated as decay products of pions, kaons, charm mesons and other hadrons produced in the far-forward direction at the LHC interaction points. The FASER collaboration recently announced the detection of such neutrino candidates with a $\sim$10 kg pilot detector~\cite{FASER:2021mtu}. Three new far-forward experiments will start their operation in upcoming Run~3 of the LHC, starting in 2022: FASER~\cite{Feng:2017uoz, FASER:2018ceo, FASER:2018bac, FASER:2018eoc}, FASER$\nu$~\cite{FASER:2020gpr, FASER:2021mtu}, and SND@LHC~\cite{SHiP:2020sos, Ahdida:2750060} and  will together collect $\sim 10, 000$ neutrino-interaction events, performing the first neutrino-interaction cross-section measurements at these energies.

To further exploit this physics opportunity, for the High-Luminosity LHC (HL-LHC) era, several ten-tonne-scale experiments are proposed. These experiments: FLArE, FASER$\nu$2, and AdvSND are being planned at a dedicated proposed Forward Physics Facility (FPF)~\cite{Anchordoqui:2021ghd, Feng:2022inv} and will detect $\sim 10^5$ $\nu_e$, $\sim 10^6$ $\nu_{\mu}$, and $\sim 10^3$ $\nu_{\tau}$ interaction events. These neutrino interaction events will significantly extend accelerator cross-section measurements by measuring neutrino interaction cross-sections in an energy range that has not been directly probed for any neutrino flavor. The large majority of neutrino interaction events are expected to fall in the DIS region, however significant number of events are also expected in the currently unconstrained SIS-DIS and soft-DIS region. In addition, by measuring the charge of the outgoing muons in charged-current interactions, muon and tau neutrinos and antineutrinos will be distinguished. These will provide the first opportunity for detailed studies of interactions of tau-neutrinos and anti-tau-neutrinos. 

\subsubsection{nuSTORM }



The Neutrinos from Stored Muons, nuSTORM, is designed to deliver a comprehensive neutrino-nucleus scattering program involving beams of $\nu_\mu$ and $\nu_e$ produced by decaying muons in a storage ring. The storage-ring instrumentation will provide neutrino flux measurements with a $1$\% or better precision~\cite{ref:nustorm}. This facility is also unique in that it can store muons with a very narrow momentum spread, which enables exact knowledge of the flavor composition of the beam and the neutrino energy spectrum. The muon beam will have momentum ranging from $1$ GeV/$c$ to $6$ GeV/$c$ with a momentum spread of $15$\%~\cite{ref:nustorm}. This will enable neutrino beams with energies ranging from $500$ MeV to roughly $5$ GeV which is similar to the kinematic range of interest in DUNE and HK~\cite{ref:nustorm}. The nuSTORM facility will also employ sophisticated neutrino-detector technologies such as those being developed for the DUNE and HK near-detector facilities. This will allow for the measurements of $\nu_{\mu}$ and $\nu_{e}$ scattering cross-sections to be performed with a percent-level precision.




\subsubsection{Polarized targets}
\label{sec:polarized_targets}

Measuring the asymmetries of neutrino cross sections on polarized nuclear targets can provide complementary access to the axial structure of nucleons and nuclei~\cite{ref:hydrogenwp}. Polarization observables can also help disentagle the nucleon pseudoscalar form factor. Due to the parity-violating nature of the weak interaction and the expected maximal polarization of the neutrino beam, these asymmetries are expected to be very large. Dynamical nuclear polarization has been employed to make polarized targets for beams of charged particles and photons, but it would have to be scaled up significantly to make a practical neutrino target. Furthermore, the particle detectors must be integrated in with the target due to the low energy of reaction products and the necessary presence of a strong magnetic field~\cite{ref:hydrogenwp}. Studies are underway to estimate the required target size to produce sufficient interactions.

\section{Measurements planned for LE program}



\subsection{CEvNS}

Many efforts are underway to measure CEvNS at existing stopped-pion and reactor facilities.
Future stopped-pion facilities are under consideration and may expand these opportunities.
At this time, positive measurements of CEvNS have been made on only two systems, CsI~\cite{Akimov:2017ade, Akimov:2021dab} and argon~\cite{COHERENT:2020iec}, by the COHERENT collaboration at the SNS stopped-pion source.
The background and threshold challenges at
reactors, where neutrino fluxes are much larger, but recoil energies are an order of magnitude lower, have yet to be overcome. These global efforts are detailed in the Snowmass whitepaper ``Coherent elastic neutrino-nucleus scattering: Terrestrial and astrophysical applications''.

\begin{table}[h]
    \centering
    \begin{tabular}{|c|c|c|} \hline
         Experiment & Source & Target \\ \hline
         COHERENT & $\pi$DAR & Na, Ar, Ge, CsI \\ \hline
         Coherent CAPTAIN Mills & $\pi$DAR & Ar \\ \hline
         JSNS$^2$ & $\pi$DAR &  \\ \hline
         ESS & $\pi$DAR &  \\ \hline
         CHILLAX & Reactor & Ar  \\ \hline
         CONNIE & Reactor & Si  \\ \hline
         CONUS & Reactor & Ge  \\ \hline
         MINER & Reactor & Ge, Si  \\ \hline
         NEON & Reactor & Na  \\ \hline
         NUCLEUS & Reactor &   \\ \hline
         NUXE & Reactor & Xe \\ \hline
         PALEOCCENE & Paleo &  \\ \hline
         Ricochet & Reactor & Ge, Zn \\ \hline
         RED-100 & Reactor & Xe \\ \hline
         NuGen & Reactor &  \\ \hline
         SBC & Reactor & Ar \\ \hline
         TEXONO & Reactor & Ge \\ \hline
         NEWSG & Reactor & H, He, C, Ne \\ \hline
    \end{tabular}
    \caption{Summary of experiments to measure CEvNS interactions.}
    \label{tab:cevnsexp}
\end{table}

\subsubsection{Stopped-Pion}
Stopped-pion facilities provide an excellent source of neutrinos in the 1 to 50 MeV range resulting from the decay of the $\pi^{+}$ to three flavors of neutrinos/antineutrinos. Since the CEvNS process does not provide the incident neutrino energy from the reconstruction of the recoil energy alone (i.e. no recoil direction), all measured cross-sections are averaged over the neutrino energy. The prompt decay of the pion to the $\nu_{\mu}$ presents the cleanest opportunity to measure the recoil distribution of a single energy. Furthermore, the difference of the pion lifetime (26~ns) and the muon lifetime (2.2~$\mu$s) in the subsequent decay to the $\bar{\nu}_{\mu}$ and the $\nu_{e}$ allows for the separation of these flavor contributions to the flux-averaged recoil distributions if the beam pulse is sufficiently short and the intrinsic timing resolution of the detector is sufficiently small compared to the muon lifetime.

Currently there are two measurements of the CEvNS interaction both made by the COHERENT collaboration at the Spallation Neutron Source at Oak Ridge National Laboratory, the first on CsI in 2017~\cite{Akimov:2017ade, Akimov:2021dab} and the second on Ar in 2020~\cite{COHERENT:2020iec}. The COHERENT collaboration is now deploying Ge and Na targets~\cite{Akimov:2018ghi} ahead of the planned SNS proton-power upgrade from 1.4 MW to 2.0 MW in 2023. Additionally, there are plans to upgrade the argon detector to increase the active mass by a factor of 25 while maintaining or improving the CEvNS sensitive threshold.
The Coherent CAPTAIN Mills collaboration has deployed a 10-ton scale single phase liquid argon detector at the Lujan Center at Los Alamos National Laboratory.

The PIP-II superconducting RF linac~\cite{pip2-2013} is currently under construction at Fermilab and is expected to be completed by the end of 2028. PIP-II is capable of operating in a continuous-wave mode and can concurrently supply 800 MeV protons to a fixed target facility in addition to LBNF/DUNE to produce a stopped-pion neutrino source. Designs for proton accumulator rings are being studied to bunch the PIP-II protons into the short pulses needed for neutrino cross-section measurements such as CEvNS. PIP2-BD is a proposed 100-ton LAr scintillation-only experiment whose detector design is inspired by the COHERENT liquid argon program and Coherent CAPTAIN Mills~\cite{pip2wp}.



\subsubsection{Reactor Measurements}
Nuclear reactors are excellent sources of electron anti-neutrinos from the  beta decays of fission products. As compared to stopped-pion sources, the neutrino fluxes are much more intense averaged over time presenting the opportunities for high statistics measurements.
This advantage comes with two significant challenges. First, the neutrino flux and energy spectrum is much more difficult to predict from theory. However, the flux above $\sim$ 2 MeV has been measured independently via the IBD process and future experiments could achieve measurements at the precision of all theoretical uncertainties. The second challenge comes from the reduced average energy of the neutrinos which produce much less energetic nuclear recoils that are either below the thresholds of existing detectors or in a region of interest dominated by background contributions~\cite{Colaresi:2022obx}. The majority of the experiments listed in Table~\ref{tab:cevnsexp} are in fact being pursued at commercial or research reactors. Nevertheless, at the time of this report there are no conclusive measurements to indicate that these challenges have been overcome, nor is an observation of CEvNS at a reactor been demonstrated.


\subsection{Inelastic Neutrino Reactions}

Inelastic interactions of sub 100 MeV neutrinos on light to heavy nuclei are a fundamental detection mechanism for oscillation experiments, sterile neutrino searches and supernova detection. For charged current events where a large and predictable fraction of the energy is transferred to the outgoing lepton, the energy of the incident neutrino can be reconstructed. Many reactor experiments detect electron antineutrinos through the inverse beta decay process on hydrogen. 
For electron neutrinos at energies above threshold, deuterium, argon, oxygen, and carbon provide scalable detection targets. There is significant community interest in dedicated programs to measure cross sections for some nuclei (e.g., argon, as summarized in the Snowmass~21 white paper ``Low-Energy Physics in Neutrino LArTPCs''~\cite{Caratelli:2022llt}), but most are only in the very early stages. An example is COHERENT's planned charged-current measurement on oxygen which will only be measured as a background to the deuteron charged-current signal. Any neutral current measurements would have to be enabled by future (and currently undefined) detector upgrades.

The IBD cross-section is well predicted at the $\sim$ 1\% level, although it is unclear if future precision experiments of reactor neutrino fluxes will find this a significant systematic uncertainty~\cite{ornlwp}. 
For charged current interactions on the deuteron, the ~3\% theoretical prediction precision  will soon become the dominant systematic for experiments utilizing this process as a normalization for stopped-pion neutrino fluxes in heavy water Cherenkov detectors~\cite{COHERENT:2021xhx} and no strategy, theoretical or experimental, to reduce this uncertainty has been developed. For larger nuclei the charged-current cross-sections predictions are less certain and in many pertinent cases have never been measured.

Inelastic neutral-current reactions are a potential source of background for precision CEvNS measurements, both as a source of neutrino-induced neutrons and in cases where the nuclear de-excitation products are undetected. Together with CEvNS and neutrino-electron elastic scattering, they also provide potential sensitivity to low-energy muon- and tau-flavor neutrinos produced by astrophysical sources.

Realizing the full sensitivities of GeV-scale accelerator neutrino oscillation experiments based on argon will require a better understanding of interaction final states in low-energy signatures. These needs are detailed in the LEPLAr WP~\cite{Caratelli:2022llt} and include a need for both improved theory and experimental measurements using pion decay-at-rest sources as well as charged particle and neutron test beams to evaluate interaction cross-sections below 100 MeV. These should be accompanied with a full evaluation of measured final states to complete and improve the modeling in generator software packages for both low energy interactions such as MARLEY~\cite{Gardiner:2021qfr,marleyPRC} and existing high energy tools such as GENIE~\cite{GENIEv3Highlights,genie}, FLUKA~\cite{Ferrari:2005zk}, and NuWro~\cite{Zmuda:2015twa}.



Low energy inelastic neutrino cross-section measurements are planned by the Coherent CAPTAIN Mills collaboration, and should achieve a 15\% uncertainty on the electron neutrino charged-current cross section after three years of running. A heavy water detector is under construction to use the charged-current interaction of the electron neutrino on deuterium, to calibrate the neutrino flux with a precision comparable to the calculated theoretical uncertainty for that process ($\sim$ 3\%)\cite{COHERENT:2021xhx,coherent:wp}. Upgrades of this detector could add sensitivity to neutral current reactions. Inelastic-interaction measurements are underway at COHERENT, and will measure neutrino-induced neutron production on Pb, Fe, and Th nuclei.
During COHERENT's argon detector operation for CEvNS, tens of charged-current events are expected and a focused analysis of that recorded data is underway. The planned upgrade to the argon detector will improve statistics with both a larger target mass and dynamic range.


Measurements of low-energy inelastic neutrino-nucleus cross-sections and related phenomena will provide helpful constraints on predictions of nuclear matrix elements for neutrinoless double beta decay~\cite{Volpe2005,Ejiri2019}.

\subsection{Parity-Violating Electron Scattering Measurements}

Parity-violating electron scattering (PVES) experiments are designed to measure the parity violating asymmetry $A_{PV}$, which is defined as the fractional difference in cross-section for positive and negative helicity electrons. Using the Born approximation, this asymmetry is proportional to the weak form factor $F_W(q^2)$ making PVES measurements useful for further understanding of CEvNS measurements. Measurements of PVES and CEvNS at the same energies are not likely in the near future due to the fundamental differences in the conditions for each process. CEvNS requires a low threshold in order to detect low-energy nuclear recoils while PVES requires the opposite in order to reduce contamination of the elastic asymmetry by inelastic contributions from excited states. These measurements provide input to modern density functional theory calculations which predict the weak form factor as measured by CEvNS experiments at low momentum transfer. Table~\ref{tab:parityviolation} shows the current PVES experimental landscape.

\begin{table}[tb]
\centering
\begin{tabular}{ c | c c c c }
\hline
 Experiment    & Target& $q^2$ (GeV$^2$)  & $A_{pv}$ (ppm)     & $\pm\delta R_n$ (\%) \\
\hline
PREX & $^{208}$Pb & 0.00616 & $0.550\pm0.018$ & 1.3 \\
CREX & $^{48}$Ca & 0.0297 & & 0.7\\
Qweak & $^{27}$Al & 0.0236 & $2.16\pm 0.19$ & 4\\
MREX  & $^{208}$Pb & 0.0073 & & 0.52\\
\hline
\end{tabular}
\caption{\label{tab:parityviolation}Parity violating elastic electron scattering experiments. Table taken from the white paper Ref.~\cite{Ankowski:2022thw}}
\end{table}




\section{Neutrino event generators}

High-quality simulations of neutrino interactions are an indispensable
requirement for successfully achieving the physics goals envisioned by the
Neutrino Frontier in the current Snowmass process. These simulations are
implemented using Monte Carlo methods in the form of computer programs called
\textit{event generators}. While the value of cross-section measurements for
refining theoretical understanding must not be understated, improvements to
neutrino event generators are the only means through which this refined
understanding can ultimately meet the interaction modeling needs of precision
experiments. In light of the critical role that event generators will play in
enabling future discoveries with neutrinos, we call for greater attention,
scrutiny, and support for their further development. Failure to invest properly
in this area will needlessly and inevitably compromise the scientific impact of
the experimental measurements (and related theoretical work) considered
throughout the other sections of this document.

Despite their status as an essential tool for the design, execution, and
interpretation of experimental analyses, neutrino event generator development
has historically been undersupported. As neutrino physics moves into the
precision era, a lack of resources for generator-related tasks will become
increasingly problematic without new investment.
There are poor incentives for those working on generators. Theory groups do not have the training nor any incentive to
 directly interface with the generator software; they spend time rightfully developing models. Experimental groups view this work as service. As it takes significant time to engage in generator implementation issues, this may fall outside the time horizon for a student or postdoc, who also must produce physics results. Furthermore, there are a limited number of experts for a given
generator, and this makes it difficult to train new people who continually move onto other work.
A particular difficulty for
securing support for neutrino event generator development is the cross-cutting
nature of the work along multiple axes. High-energy physics provides many of
the scientific questions which can motivate generator improvements, but much of
the necessary expertise belongs to the nuclear physics community. 
Attracting and retaining sufficient scientific talent
to meet the generator-related needs of the future neutrino program will likely
require establishment of \textit{computational neutrino interaction physics} as
a viable research emphasis and career path.  

Support from laboratories will likely be critical for long-term event generator
infrastructure and for facilitating the necessary collaboration across the
entire neutrino community. We look to the LHC and the GEANT
effort~\cite{hepgenWhitePaper} for successful models of engagement and career
progression. The major neutrino experiments themselves are also an important
stakeholder for future event generator work. 
For example, DUNE, like the LHC, will need multiple generators which not only meet the needs of the experiment but also interface to the experiment in a productive way. Therefore, it is helpful if DUNE can consider and decide upon a potential model for work to proceed. This may lead to interface or other technical requirements on generators not covered here. Discussion and articulation of the interface of generators with experiments could be fruitful.

While issues related to software and computing are obviously important for
neutrino event generators, it should be emphasized that the challenges faced by
the community are not purely technical. Experiments require the full final
state of every neutrino interaction to be simulated, despite many gaps in the
current theoretical understanding of the relevant nuclear physics. Considerable
scientific judgment is therefore required in selecting the patchwork of models
to be implemented, accounting for their ranges of validity appropriately,
tuning free parameters to cross-section data, and thoroughly quantifying
systematic uncertainties. Close collaboration between theorists,
experimentalists, and generator developers on these topics is optimal, and some
success stories exist, including a series of invited speakers at internal T2K
meetings, the Neutrino Theory Network, and a joint theory-experiment working group at
Fermilab~\cite{FNALTheoryExperiment}. The Neutrino Scattering Theory Experimental Collaboration includes generator, theory and experimental representation.  Ideally, all of these activities are
guided by comparisons to neutrino-, electron-, pion-, and
photon-nucleus\footnote{There is a lot of ($\gamma$,A) data with photon energy
around 1 GeV, especially from bremsstrahlung (e.g., the TAPS experiment). As
neutrino cross sections probe small $Q^2$,  photon-cross section at $Q^2=0$
provide a useful benchmark. GiBUU includes photoproduction data in its overall
validation.} scattering data.\footnote{Tools to enable these comparisons are also dedicated software products which are far from trivial to implement. A widely-used example is NUISANCE~\cite{Stowell:2016jfr}.} In some cases (e.g., neutron emission from
tens-of-MeV charged-current inelastic $\nu_e$-argon
scattering~\cite{marleyPRC}), an event generator currently provides the only
theoretical prediction available for a particular scattering process.

High-quality event generators are especially crucial for enabling a broad program of searches for physics beyond the Standard Model in the neutrino sector. For many scenarios, the near detector constraints used in oscillation analyses will be of limited utility, and thus both standard physics and a potential exotic signal will need to be simulated with high precision. Incentives should thoughtfully be put into place to support the full range of needed model-building and generator development activities.

Many needs and challenges for neutrino generators are shared by event generator efforts in other areas of high-energy physics. A dedicated Snowmass white paper~\cite{hepgenWhitePaper} provides a broad overview of the field as a whole. Experience and techniques from, e.g., the LHC community may usefully be applied to neutrino interaction simulations and should be further explored.

\subsection{Low energies}

Nearly all development activity for neutrino event generators has been focused
to date on the intermediate energies ($\sim$0.1--20 GeV) relevant for
accelerator-based oscillation experiments. This emphasis is well-motivated, but
strategic planning discussions for the field should be mindful of the
interaction simulation needs of the broader neutrino community as well. At low
energies ($\lesssim$100 MeV), precise simulations for multiple important
interaction modes are relatively straightforward to implement apart from
obtaining sufficiently high-quality theoretical inputs, such as the neutron
form factors for CEvNS. The major exception is low-energy inelastic scattering
on complex nuclei, which remains poorly understood theoretically and is too
sensitive to nuclear structure details for the prevailing treatments used by
event generators at higher energies to provide a good description. A dedicated
event generator called MARLEY~\cite{marleyPRC,Gardiner:2021qfr} provides the
capability to simulate $\isotope[40]{Ar}(\nu_e,e^{-}) \isotope[40]{K}^{*}$ and
other inelastic reactions, but it is currently supported by a
single untenured individual at a tiny fraction of 1 FTE. Substantial further
development of MARLEY or a similar event generator will be required to fully
execute experimental analyses (such as DUNE supernova or solar neutrino
measurements) that rely upon an understanding of sub-100-MeV inelastic
neutrino-nucleus scattering.

\subsection{Intermediate energies}

At accelerator energies, the neutrino event generator landscape is presently
dominated by four modern codes: GENIE~\cite{genie,GENIEv3Highlights},
GiBUU~\cite{Buss:2011mx}, NEUT~\cite{Hayato:2009zz,Hayato2021}, and
NuWro~\cite{golan2012}. None of these is currently maintained by more than a
handful of active developers, nearly all of whom have significant additional
responsibilities. The major generator groups differ in their overall goals and
in the technical and scientific scope of the simulations that they seek to
provide. Both GiBUU and NuWro are developed primarily as tools for theoretical
investigation. The GiBUU effort aims to achieve the best possible unified model
for lepton-, photon-, and hadron-nucleus scattering, including collisions of
heavy ions. NuWro is more narrowly focused on neutrino and
electron~\cite{Zmuda:2015twa} interactions, and it allows simulations to be
performed using a relatively wide range of model ingredients. 
Theory-based generators can play a
critical role; NuWro is used on T2K to provide independent checks of model choices and implementation, and as a ``sandbox'' for new ideas, without the overhead of a full production.

The authors of GENIE and NEUT are mostly experimentalists, and the development
of both codes is driven by the needs of that community. In particular, both GENIE and NEUT provide extensive capabilities for essential tasks in experimental production workflows, e.g., interfacing with beam and detector simulations and systematic uncertainty quantification. Improvements
to NEUT are focused on the priorities of Super-K, T2K, and Hyper-K, which use
NEUT as their primary event generator. Despite contributions from members of
those experiments and interested theorists, ``the lack of human resources
render it difficult to support NEUT as a more general tool,''~\cite{Hayato2021}
and the code remains closed-source due to this limitation. GENIE is not tied to
any specific experiment, but it is used as the primary event generator by all
Fermilab neutrino experiments, and it seeks to be a universal and flexible
simulation platform for the field as a whole.

A successful strategy for neutrino event generator development should leave
room for innovations outside of well-established efforts, including targeted,
process-specific generators (which might interface with existing
infrastructure) as well as codes that employ new techniques to deliver a
potentially broad range of physics simulations. A prototype example of the
latter is ACHILLES~\cite{achilles}, which applies multiple methods from LHC
event generators to lepton-nucleus scattering for the first time. In
particular, the ACHILLES capability to calculate the leptonic current for
arbitrary beyond-the-Standard-Model processes~\cite{achillesMatrixElement} may
enable a wide range of exotic physics simulations without a dedicated
implementation effort on a per-model basis.

\subsection{Common challenges}

There are a number of shared challenges for the neutrino event generator
community which are widely recognized and for which meaningful progress can be
achieved with increased effort. A partial list of these challenges, some of
which were discussed at a recent series of workshops at ECT$^*$ and
Fermilab~\cite{FNALgentools}, is given below.

\begin{enumerate}

\item The addition of a new cross-section model in an event generator is
typically labor-intensive, requiring multiple person-years to fully execute and
validate. While some level of human effort and evaluation (e.g., with respect
to model suitability for different nuclear targets and kinematic regions) will
always be essential, technical strategies for shortening the implementation
timeline are worthy of further investigation. Streamlining model integration in
this way will make it easier for theory groups to directly contribute to
experimental programs.

\item Experimental analyses typically modify an a priori event generator
prediction with data-driven constraints (e.g., from a near detector in an
oscillation experiment) to improve agreement with simulation and reduce
systematic uncertainties. Tools for applying these constraints are typically
internal to an experiment and often heavily dependent on a particular event
generator configuration. These limitations prevent detailed, quantitative
studies about which generator model improvements should be prioritized to
maximize future experimental sensitivity.

\item No common standard currently exists for software interfaces between
neutrino event generators and the beam and detector simulations
used in experimental production workflows. This substantially reduces the
ability of neutrino experiments to use multiple event generators in the
design and execution of analyses.

\item There is similarly no common standard for the format of the
\textit{event record} used to store the list of particles
simulated in a single neutrino interaction. This raises similar issues
as the missing beam/detector interfaces, and it also impedes information
passing between event generators. The LHC community has substantially
benefited from a common event record format (HepMC3~\cite{Buckley2021} and
its predecessors).

\item As neutrino interaction simulations become more sophisticated, the
resource needs associated with event generation are likely to become an
important part of the overall computing budget for large experiments (as they
already are for the LHC). Research and development towards maximizing the event
generation efficiency in that context (including repurposing existing solutions
from other subfields) will be worthwhile to pursue.

\end{enumerate}

The pandemic has significantly slowed effort to resolve many of the issues
around neutrino event generators. Face-to-face time to explore solutions is
critical, especially in light of the broad group of developers and users that
should be involved in such discussions. Building upon the precedent of
LHC-related meetings at Les Houches, we suggest that a second series of
community workshops about neutrino event generators be held in the near future.
Items for discussion may include strategies for better organizing related work,
reflections on the success and shortcomings of recent efforts, and a variety of
more technical topics. 
\section{Data Preservation and Archival}

The preservation of data from modern neutrino experiments will become more important as such experiments become more complex. It is important that the neutrino community preserves the data from neutrino cross-section measurements so that future experiments have access to these important measurements. The current preservation of cross-section data from experiments such as MINERvA~\cite{Fine:2020snd} and COHERENT~\cite{coherent:wp}
leads to useful phenomenological studies and aids in the tuning of models used in oscillation experiments~\cite{datapresref,tensions2019report}.
Experiments should strive to make their results as accessible as possible. The Durham High-Energy Physics Database (HEPData)~\cite{Maguire2017,hepdata} is a valuable open-access repository for experimental data sets, including neutrino cross-section measurements. Other online data storage systems, such as Zenodo~\cite{Zenodo}, can also play an important role in enabling experimental data releases.
Experimental/institutional web pages should be structured to make data releases easily searchable.

There were several important considerations raised for preserving data from cross-section measurements. Including correlation matrices within data releases is an essential requirement for a fuller understanding of systematic uncertainties. Experiments should strive to include this information into the data release materials. Efficiency corrections or limitations of the simulations should be acknowledged within any published data releases. Including this information allows for an independent check that a user can apply to the published analysis. When an analysis applies corrections to convert measured quantities to their true values (i.e., applies an \textit{unfolding}), model comparisons should be carried out in both reconstructed and truth spaces for parameters of interest. Experiments should also consider reporting data before an unfolding takes place to expand the utility of the data to an outside user. Another consideration is that taking steps to preserve data in a way that is most useful to the community requires resources within a collaboration. Often the members of the experiment most influential to the analysis and data preservation are no longer available if an analysis is examined many years later through the preserved data.

The number and sophistication of neutrino scattering data sets will expand at an accelerating rate in the coming years. For the community to most fully benefit from these measurements, a greater level of organization and standardization is needed. A centralized \textit{neutrino scattering center} focused on curating relevant cross-section data and supporting downstream users would be a solution worthy of serious consideration. Such a center could also define standard guidelines for experimental collaborations to follow in preparing their data releases, thus helping to ensure the long-term usefulness and robustness of the results. The mandate for the center should be broad enough to encompass relevant data sets from electron and hadron beam experiments. Precedents from other areas of the discipline, including the CERN data center from the LHC community and the National Nuclear Data Center (NNDC)~\cite{nndc} for nuclear physics, can provide helpful models for what might be done for neutrino physics.


\section{Acknowledgements}
\label{sec:acknowledgements}


The authors would like to thank the collaborations which provided material for this report outside of the white papers: ANNIE, DUNE (and ProtoDUNE), HK, ICARUS, LArIAT,  MINERvA, MicroBooNE, NINJA, NOvA,  SBND, T2K, and WCTE. The authors further thank Tom Junk for drafting Sec.~\ref{sec:polarized_targets} about polarized targets. The authors gratefully acknowledge the white papers received. These are summarised in Appendix~\ref{app:wplist}.
The authors would also like to gratefully acknowledge the comments received from the community on earlier drafts of this summary document.


\appendix

\section{List of whitepapers submitted for NF06}\label{app:wplist}


\begin{itemize}

\item A.~M.~Ankowski, \textit{et al.}, ``Electron Scattering and Neutrino Physics,''
arXiv:2203.06853 [hep-ex].

\item L.~Alvarez-Ruso, \textit{et al.}, ``Theoretical tools for neutrino scattering: interplay between lattice QCD, EFTs, nuclear physics, phenomenology, and neutrino event generators,''
arXiv:2203.09030 [hep-ph].


\item L.~Alvarez-Ruso, \textit{et al.},``Neutrino Scattering Measurements on Hydrogen and Deuterium: A Snowmass White Paper,'' arXiv:2203.11298 [hep-ex].

\item L.~Alvarez-Ruso, \textit{et al.},``Bubble Chamber Detectors with Light Nuclear Targets: A Snowmass 2021 White Paper,'' arXiv:2203.11319 [physics.ins-det].

\item J.~M.~Campbell, \textit{et al.}, ``Event Generators for High-Energy Physics Experiments,'' arXiv:2203.11110 [hep-ph].


\item A.~A.~Abud \textit{et al.} [DUNE Collaboration], ``Snowmass Neutrino Frontier: DUNE Physics Summary,'' arXiv:2203.06100 [hep-ex].

\item A.~A.~Abud \textit{et al.} [DUNE Collaboration], ``A Gaseous Argon-Based Near Detector to Enhance the Physics Capabilities of DUNE,'' arXiv:2203.06281 [hep-ex].

\item D.~Caratelli, \textit{et al.}, ``Low-Energy Physics in Neutrino LArTPCs,''
arXiv:2203.00740 [physics.ins-det]. 

\item M.~Andriamirado, \textit{et al.}, ``Physics Opportunities with PROSPECT-II,''
arXiv:2202.12343 [hep-ex].

\item J.~L.~Feng, \textit{et al.}, ``The Forward Physics Facility at the High-Luminosity LHC,'' arXiv:2203.05090 [hep-ex].

\item 
C.~W.~Bauer,  \textit{et al.}
``Quantum Simulation for High Energy Physics,''
arXiv:2204.03381 [quant-ph].


\item
M.~Abdullah,  \textit{et al.}
``Coherent elastic neutrino-nucleus scattering: Terrestrial and astrophysical applications,''
arXiv:2203.07361 [hep-ph].


\end{itemize}


\renewcommand{\refname}{References}

\printglossary

\bibliographystyle{utphys}

\bibliography{common/tdr-citedb}

\end{document}